# Multiscale approach including microfibril scale to assess elastic constants of cortical bone based on neural network computation and homogenisation method.

Abdelwahed BARKAOUI[1*], Abdessalem CHAMEKH[2], Tarek MERZOUKI[3], Ridha HAMBLI[1], Ali MKADDEM[4]

[1]*PRISME laboratory, EA4229, University of Orleans*
*Polytech' Orléans, 8, Rue Léonard de Vinci 45072 Orléans, France.*

[2] *Mechanical Engineering departement, Faculty of Engineering, King Abdulaziz University,*
*P.O. Box: 80204, Jeddah 21589, Saudia arabia.*

[3]*Laboratoire Ingénierie des Systèmes de Versailles, Université de Versailles St Quentin en Yvelines*
*10 avenue de l'Europe, 78140 Velizy.*

[4]*Arts et Métiers ParisTech, LMPF-EA4106, Rue Saint Dominique BP508, 51006 Châlons-en-Champagne, France.*

## SUMMARY

The complexity and heterogeneity of bone tissue require a multiscale modelling to understand its mechanical behaviour and its remodelling mechanisms. In this paper, a novel multiscale hierarchical approach including microfibril scale based on hybrid neural network computation and homogenisation equations was developed to link nanoscopic and macroscopic scales to estimate the elastic properties of human cortical bone. The multiscale model is divided into three main phases: (i) in step 0, the elastic constants of collagen-water and mineral-water composites are calculated by averaging the upper and lower Hill bounds; (ii) in step 1, the elastic properties of the collagen microfibril are computed using a trained neural network simulation. Finite element (FE) calculation is performed at nanoscopic levels to provide a database to train an in-house neural network program; (iii) in steps 2 to 10 from fibril to continuum cortical bone tissue, homogenisation equations are used to perform the computation at the higher scales. The neural network outputs (elastic properties of the microfibril) are used as inputs for the homogenisation computation to determine the properties of mineralised collagen fibril. The mechanical and geometrical properties of bone constituents (mineral, collagen and cross-links) as well as the porosity were taken in consideration. This paper aims to predict analytically the effective elastic constants of cortical bone by modelling its elastic response at these different scales, ranging from the nanostructural to mesostructural levels. Our findings of the lowest scale's output were well integrated with the other higher levels and serve as inputs for the next higher scale modelling. Good agreement was obtained between our predicted results and literature data.

## 1. Introduction

Hierarchical structures in bio-composites such as bone tissue have many scales or levels, specific interactions between these levels and a highly complex architecture in order to fulfil their biological and mechanical functions [1, 2]. The complexity and heterogeneity of bone tissue require a multiscale modelling to understand its mechanical behaviour and its remodelling mechanism [3,4].

Katz et al. [5] distinguish five levels of hierarchical organization, shown in figure 1, which have been widely accepted in the scientific community: (i) the macrostructural level: several millimeters to several centimeters depending on the species, or whole bone level, consisting of cortical and trabecular bone types,(ii) the microstructure at an observation scale of several hundred millimeters to several millimeters, where cylindrical units called osteons build up cortical bone, (iii) the ultrastructure (or extracellular solid bone matrix) at an observation scale of several millimeters, comprising the material building up both trabecular struts and osteons. Trabecular struts or plates can be distinguished, (iv) within the ultrastructure, collagen-rich domains and collagen-free domains can be distinguished at an observation scale of several hundred nanometers. Commonly, these domains are referred to as fibrils and extrafibrillar space [6] and (V) the nano-scale: several ten of nanometers, the so-called elementary components (hydroxyapatite, collagen) of mineralised tissues can be distinguished. Feng et al. [7] distinguish five levels. Nanostructural level: ranging from few nanometers to several hundred nanometers. At this level, the bone can be considered as a multiphase nanocomposite material consisting of an organic phase (32–44% bone volume), an inorganic phase (33-43% bone volume) and water (15–25% bone volume). Sub-microstructural level: also called a single lamella level (spanning 1 to a few microns). Microstructural level: tenths to hundreds microns, or a single osteon and interstitial lamella level. Mesostructural level, several hundred microns to several millimeters is the cortical bone level. Macrostructural level: several millimeters to several centimeters, depending on the species, or whole bone level. Barkaoui and Hambli [2] provide a more detailed description of the multi-scale structure. They suggest that, in addition to the scale of the whole bone and scale of the basic components (mineral, collagen and water), human cortical bone structure consists of six structural scale levels, namely (macroscopic) cortical bone, osteon, lamella, fiber, fibril and microfibril scales. They show the importance of the microfibril scale in the mechanical behavior of bone, and also study the influence of cross-links at this scale. In fact, microscopic analysis reveals a complex architecture that can be described as hollow cylinders juxtaposed next to each other and sealed by a matrix. The cylinders are called osteons, the holes are named Haversian canals and the matrix is the interstitial system. Further analysis shows that osteons are in fact an assembly of cylindrical strips embedded in each other. Lamella is composed of a network of collagen fibers with a helical orientation and inserted into hydroxyapatite crystals. The orientation of collagen fibers may be different between two consecutive lamellae. Every fiber is formed by a set of fibrils. Each fibril is in turn composed of microfibrils. Finally, each microfibril is a helical arrangement of five tropocollagen (TC) molecules.

Many researchers have recently addressed this problem by developing analytical or numerical models of the elastic behavior of cortical bone [8-14] and fracture behaviour [6, 15-17].These models use homogenisation techniques, some of them with a multiscale approach, but none gives a complete description of the hierarchical organization of cortical bone. Some studies focus only on one level scale of: lamellas [18], osteons [19], or fibrils [20]. The multiscale hierarchical organization is, however, addressed in the work of Martínez-Reina et al., Hamed et al. 2010 and Fritsch and Hellmich [6, 21, 22]. All of them use a multiscale approach, starting at the nano scale, from the basic constituents of bone tissue: mineral, collagen, water, and going up to the micro scale.

The use and generalization of the above models to include changes in the mechanical and geometrical parameters is not simple. In this work, the multiscale approach is carried out in three phases. Phase 0 includes the Hill bounds equation to determine the elastic constants of collagen-water and mineral-water composites. Phase 1 concerns the microfibril level. In this second phase, the Neural Network (NN) method was used to calculate the elastic properties of mineralised collagen microfibrils. This method also makes it possible to vary the geometrical and mechanical parameters of the elementary constituents and observe their influence on the properties, which is difficult using the homogenisation equations. The NN is trained by using data from a 3D finite element study of the microfibril levels [2]. Phase 2, from fibril to continuum cortical bone tissue, homogenisation equations [18] are used to perform the computation at the higher scales. The elastic properties of microfibril obtained by the neural network are used as inputs for the homogenisation computation to determine the properties of fibril.

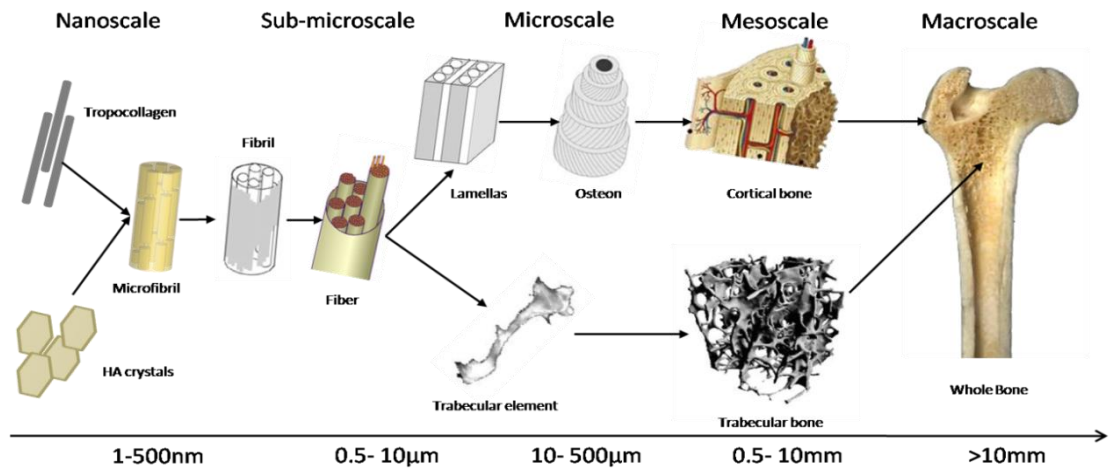

**Figure 1.** Multiscale Hierarchical structure of bone

The main novelties of the current method are: (i) the implementation of the lowest scale of bone structure by explicitly considering the nanoscale basic constituents (mineral volume fraction and properties, collagen state, mechanical properties and the number of cross-links); (ii) implementation of a multiscale hybrid method (NN for low scales and homogenisation for higher scales) to provide material properties to simulate bone behaviour (fracture, remodelling). This multiscale approach is based on three methods: finite elements (FE), NN and homogenisation, enables investigation of the structure-property relationships in human bone. The advantage of this method that it can be used to define the equivalents properties for a class of parameterized unit cells. Moreover, it can be easily incorporated in the finite element code to model the entire bone structure. The originality of this work is to use the Neural Network (NN) method to calculate the elastic properties of mineralized collagen microfibrils. The later does not used in other works and the microfibril level is not considered between bone elementary constituents (collagen and mineral) and mineralized collagen fibril levels. The Neural Network (NN) method allows to change the geometrical and mechanical parameters of the elementary constituents and observe their influence on the properties, which is difficult using the homogenization equations. The elastic properties of microfibril obtained by the neural network are then used as inputs for the homogenization computation to determine the properties of fibril.

## 2. Materials and methods

In this section, a demonstration of the proposed multiscale approach in addition to the elementary components of the bone and their Hill bounds will be cited. A description of the proposed model steps along with the methods which are used in this approach will be introduced. In the results and discussion section a validation of this approach will be adduced once the predicted elastic modulus will be compared with the available pervious numerical studies and experimental data.

### 2.1. Bone composition

The elementary components of bone can be distinguished as follows:

**Collagen**: At the lowest hierarchical level, bone structure is composed of collagen molecules which can be viewed as rods of about 300 nm long and 1.5 nm in diameter, made up of three polypeptide strands, each of which is a left-handed helix [23]. The mineralization process reduces the lateral spacing between collagen molecules bringing them closer to one another [24]. The arrangement of the TC molecules comes from the strong chemical bonds (cross-linking) that are formed between adjacent collagen molecules throughout the collagen bundles [25,26].

**Mineral**: The mineral phase is almost entirely composed of impure hydroxyapatite crystals [$Ca_{10}(PO4)_6(OH)_2$]. Those crystals are plate-like shapes. The size of the mineral plates varies according to the bone type (cortical or trabecular, human or animal, etc.) [27,28]. A wide range of mineral plate dimensions has been reported in the literature: 15–150 nm in length, 10–80 nm in width and 2–7 nm in thickness, while the distance between the neighboring plates is on the same order as the thickness [29]. During the bone remodeling process, mineral particles are nucleated primarily inside the gap region of the microfibrils. In a later stage, the mineral particles extend into the extrafibrillar region [30,31]. Hydroxyapatite mineral is stiff and extremely fragile, exhibiting elastic isotropic behavior [32, 33]. In this study we assume that the mineral phase occupies all the space between the TC molecules, because the mineral component is essentially crystalline, but may be present in amorphous forms [33-38].

**Cross-links**: The fiber structure of bone is stabilized through intermolecular cross-links joining two collagen molecules. Collagen cross-linking plays a critical role in the connectivity of bone microfibrils, fibrils and fibers [39-43]. Cross linking is either enzymatically or non-enzymatically mediated [44]. It has been reported that the formation of intermolecular covalent cross-links has a significant effect on material properties (strength and brittleness) [45-50] and mechanical behavior [2, 44, 51-54].

**Water**: Water is the third major component in bone and occupies about 10–25% of the whole bone mass [55]. It provides the liquid environment for the biochemical activity of the non-collagenous organic matter. Recent theoretical and experimental work has highlighted the important role of water in the failure properties of bone by a mechanism which consists in gluing together the collagen and mineral phases [14, 56, 57]. At the ultrastructure level of bone, water exists in the form of bound water in the collagen network (including the collagen-mineral interface) and tightly bound water in the mineral phase [58]. Consequently, water may play an important role in bone behavior involving creep, fatigue and fracture [59-62].

### 2.2. Neural network method description

The NN model is a parallel processing architecture consisting of a large number of inter-connected processing elements called neurons organized in layers. A NN model can be used to map input to output data without a known 'a priori' relationship between the data sets. NNs have been widely and increasingly employed in the analysis of problems in science and technology [63-66]. One of the distinct characteristics of the NN is its ability to learn and generalize from experience and examples and to adapt to changing situations following an initial training phase. NNs are able to map causal models (i.e. mapping from cause to effect for estimation and prediction) and to conduct inverse mapping (i.e. mapping from effect to possible cause) [64].

The single neuron performs a weighted sum of the inputs $x_i$ that are generally the outputs of the neurons of the previous layer $v_m$, adding threshold value and producing an output given by:

$$v_m = \sum_{i=1}^{L} w_{im} \, x_i + b_i \quad (1)$$

$w_{im}$ are the network weights.

The training process in the NN involves presenting a set of examples (input patterns) with known outputs (target output). The system adjusts the weights of the internal connections to minimize errors between the network output and target output. The knowledge is represented and stored by the strength (weights) of the connections between the neurons [63-66]. Once the

NN has been satisfactorily trained and tested, it is able to generalize rules and respond to input data in order to predict the required output rapidly (in seconds) within the domain covered by the training examples [63-66]. There are several algorithms in a NN and the one used in the current analysis is the BP training algorithm. The BP algorithm is an iterative gradient algorithm designed to compute the connection weights, minimizing the total mean square error between the actual output of the multilayer network and the desired output. In particular, the weights are initially chosen randomly and the rule consists of a comparison of the known and desired output value with the calculating output value by utilizing the current set of weights and threshold.

The mean square error is calculated by:

$$J = \left(\frac{1}{2}\frac{1}{P}\right)\sum_{1}^{P}\sum_{i-1}^{N}(D_{im} - \gamma_{im})^2 \quad (2)$$

where $\gamma_{im}$ is the actual output of the *i*th output node with regard to the *m*th training pattern, while $D_{im}$ is the corresponding desired output. P and N denote respectively the total number of patterns and the number of output nodes,

### 2.3. Multiscale model description

In this study, an algorithm composed of 11 steps is employed to estimate the elastic properties (Figure 2). The three main elementary components of bone material: collagen, hydroxyapatite and water, are grouped into two composites, namely collagen-water and mineral-water. Considering that the water embedded in the space between collagen and hydroxyapatite is bound to both phases. Step 0 consists to determine the elastic constants of collagen-water and mineral-water composites by averaging the upper and lower Hill bounds. With the cross-links, these two composites form collagen microfibrils, whose elastic constants are calculated in Step 1 using the FE method and are generalized using the NN method. The method of Nemat-Nasser and Hori [67] for composites with periodically distributed inclusions was applied in Steps 2, 3 and 4 to calculate the stiffness tensor of collagen fibrils, fibers (composed of fibrils surrounded by mineral-water) and lamellae (fibers surrounded by mineral-water). In Step 5, the canaliculi porosity is taken into account, and the tensor stiffness of the lamellae with canaliculi is also calculated using the method of Nemat-Nasser and Hori [67]. Canaliculi have been simplified as straight tubes, and are oriented in three directions within the osteon: radial, longitudinal and circumferential. In order to take each type of canaliculi in to account, three different laminate structures were considered. In Step 6, the stiffness tensor of a material comprising the three types of canaliculi was obtained through a rule of mixture specific for laminates and developed by Chou et al. 1972 [68]. In Step 7, lacunar porosity is introduced in the model in the same way as canalicular porosity in step 5, just changing the pore shape. In Step 8, given the approximately cylindrical shape of osteons, a symmetrization technique is used to assess thee effective transversely isotropic elastic constants of a single osteon. In Step 9, osteons with different mineral content (even interstitial bone, simplified here as osteons with a very high mineral content) are taken as phases that interpenetrate each other to form a new composite. The stiffness tensor of this composite is estimated using a self-consistent scheme, by taking into account the superimposition of osteons in the cortical tissue as a consequence of bone remodeling. In Step 10, vascular porosity is included in the model by means of a Mori-Tanaka scheme with diluted inclusions representing the Haversian canals.

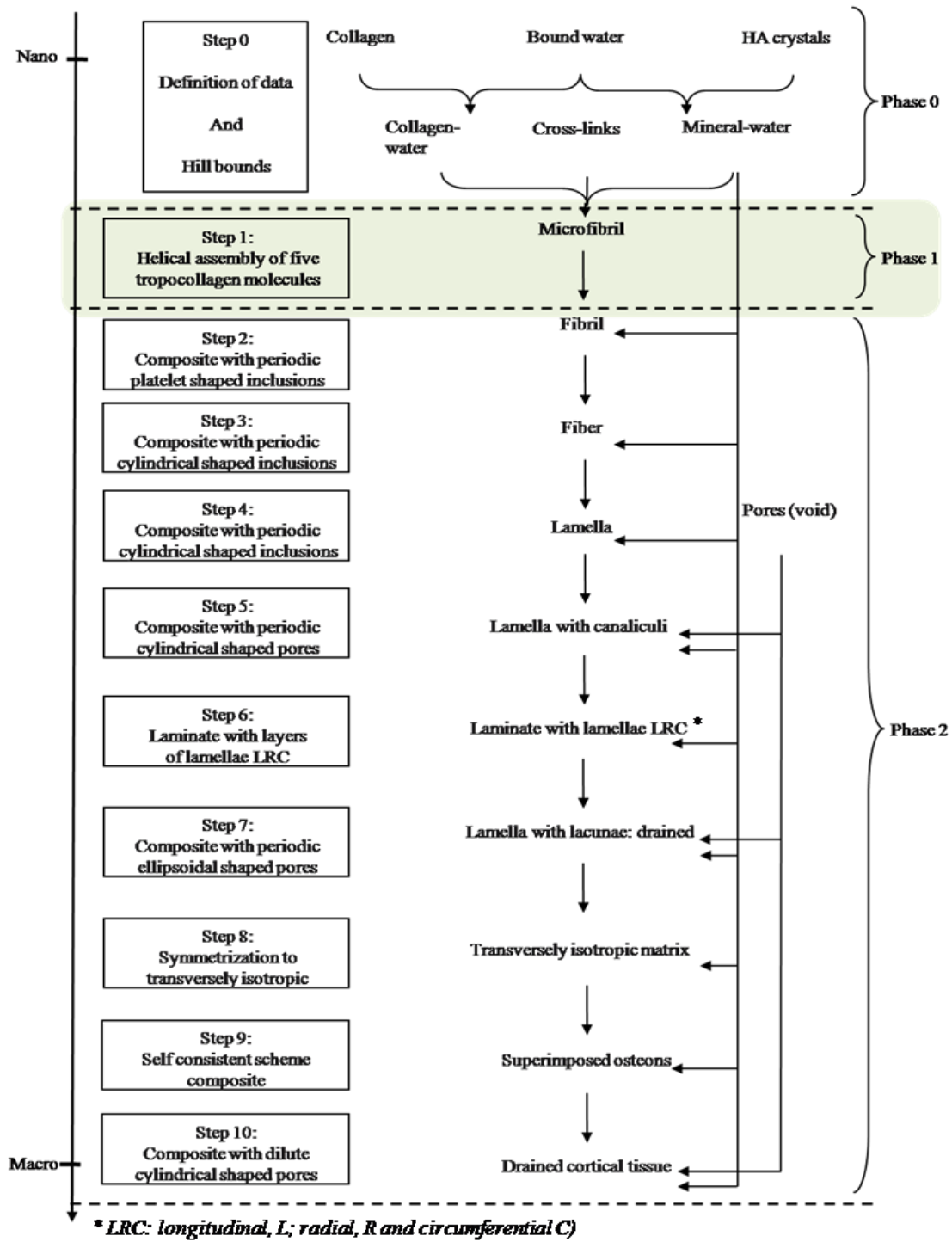

**Figure 2.** Algorithm employed for the estimation of the elastic properties of cortical tissue.

### 2.4. Multiscale modeling of cortical bone

The multi scale approach of the current study comprises the following steps, going from step (0) which represents the Hill bounds to step (10) which represents the cortical bone.

### 2.4.1. Step 0: Hill bounds

In the present work, the elastic constants of the collagen and hydroxyapatite minerals are assumed to be isotropic based on the explanation of Yoon and Cowin [18, 19]. Yoon and Cowin [18, 19] consider that the elastic constants of the hydroxyapatite minerals are assumed to be isotropic because the anisotropic elastic constants of the hydroxyapatite minerals in the human bone are not yet reported even though the hydroxyapatite crystals are known to have the hexagonal symmetry. Morin et al., [69] consider that the collagen matrix anisotropic or approximately transversely isotropic. The elastic constant values employed for collagen and hydroxyapatite are taken from Gong et al. [70] and Biltz and Pellegrino [71]. Hill bounds, which are tensorial forms of the classical Voigt (V) and Reuss (R) bounds, are estimated for the collagen-water and mineral-water composites using the following equations [18]:

$$C_{cw}^V = (1 - \phi_{wc})C_c + \phi_{wc}C_w \tag{3}$$

$$C_{hw}^V = (1 - \phi_{wh})C_h + \phi_{wh}C_w \tag{4}$$

$$C_{hw}' = (1 - \phi'_{wh})C_h + \phi'_{wh}C_w \tag{5}$$

$$E_{cw}^V = (1 - \phi_{wc})E_c + \phi_{wc}E_w \tag{6}$$

$$E_{hw}^V = (1 - \phi_{wh})E_h + \phi_{wh}E_w \tag{7}$$

Where C and E designate stiffness and compliance tensors, respectively, and $\phi_{wc}$ and $\phi_{wh}$ are the occupied volumetric fraction of water in the collagen-water and mineral-water composites, respectively. The shear modulus $\mu$ and the bulk modulus K for the collagen-water (cw) and mineral-water (hw) of the classical Voigt (V) and Reuss (R) bounds are estimated from the previous isotropic tensors. The average of the two bounds Voigt (V) and Reuss (R) is used to estimate the shear and bulk modulus of the collagen-water and mineral-water [18]:

$$\begin{cases} \mu_{cw} = \dfrac{\mu_{cw}^V + \mu_{cw}^R}{2} \\ K_{cw} = \dfrac{K_{cw}^V + K_{cw}^R}{2} \end{cases} \tag{8}$$

$$\begin{cases} \mu_{hw} = \dfrac{\mu_{hw}^V + \mu_{hw}^R}{2} \\ K_{hw} = \dfrac{K_{hw}^V + K_{hw}^R}{2} \end{cases} \tag{9}$$

These expressions have been extensively used to estimate isotropic elastic constants of polycrystals. The isotropic tensors of the collagen-water and mineral-water $C_{cw}$ and $C_{hw}$ are then deduced.

### 2.4.2. Step1: Mineralised collagen microfibril

This section describes how the mineralised collagen microfibril was modeled i.e. (i) the 3D FE model proposed by Barkaoui and Hambli, Hambli and Barkaoui [2,54,72] was employed in this step, (ii) the NN method was introduced to obtain a general statement on the behavior of mineralised collagen microfibrils and (iii) mechanical elastic properties and their effects on the elastic behavior of microfibrils were assigned to be the data base for the NN training. This step presents two principal stages: the 3D finite element model and the trained neural network.

**a) 3D finite element modeling**

The microfibril is a helical assembly of five TC molecules (rotational symmetry of order 5), which are offset one another with an apparent periodicity of 67 nm (Figure 2). This periodical length is denoted by the letter D and is used as a primary reference scale in describing the structural levels. The helical length of a collagen molecule is 4.34 D ≈ 291 nm and the discrete gap (hole zone) is 0.66 D ≈ 44 nm between two consecutivetype1 TC molecules in a strand. These gaps in bone are the sites of nucleation for hydroxyapatite crystals (the mineral component of bone tissue) to be deposited. The five molecules create a cylindrical formation with a diameter 3.5–4 nm and of unknown length [2,54]. The orientation and axial arrangement of TC molecules in the microfibrils were deduced from electron microscopic observations showing transverse striations with a period D. These striations are caused by the staggered arrangement of TC molecules at an interval D.  The strength and stability of the microfibrils during maturation are achieved by the development of intermolecular cross-links.

Barkaoui and Hambli [2, 54,72] used a three dimensional model with a cylindrical form [73] of the mineralised collagen microfibril to perform a FE analysis using the ABAQUS software. The 3D geometry model is meshed using tetrahedral elements (Figure 3) and solved via the ABAQUS standard scheme. For the boundary conditions, it is assumed that the mineralised collagen microfibril retains its cylindrical form during elastic loading. The left surface of the microfibril was encastred and a uniaxial force (F) along the axis of the collagen molecules was applied to the right surface of the microfibril. Deformation and elongation Δl were computed. Plasticity and rupture in both phases (collagen and mineral) are beyond the scope of the current work. It has been reported that relative sliding on the interface between two phases plays a role in the fracturing process of the fibrils [51, 74]. Sliding effects are therefore neglected here for simplicity.

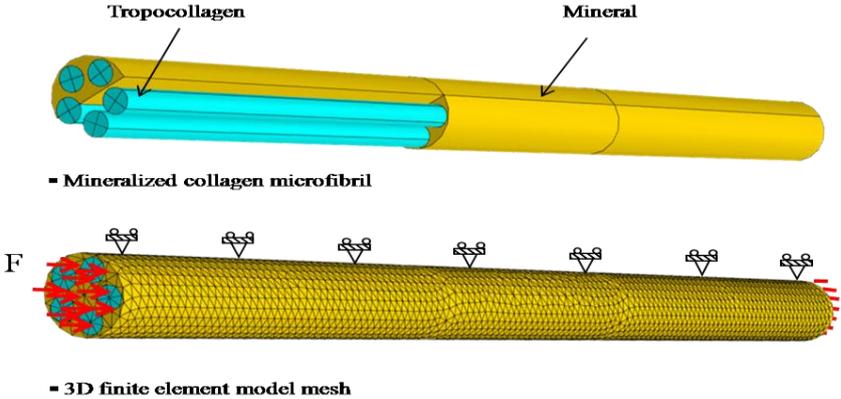

Figur3. Three-dimensional finite element model of mineralized collagen microfibril, with boundary conditions of the compression simulation [2, 54].

The results of the 3D FE simulation of collagen microfibril found by Barkaoui and Hambli [2] are shown in Figures (4) and (5). These results were considered to be the data base for the NN training in order to have a generalized statement for the following level.

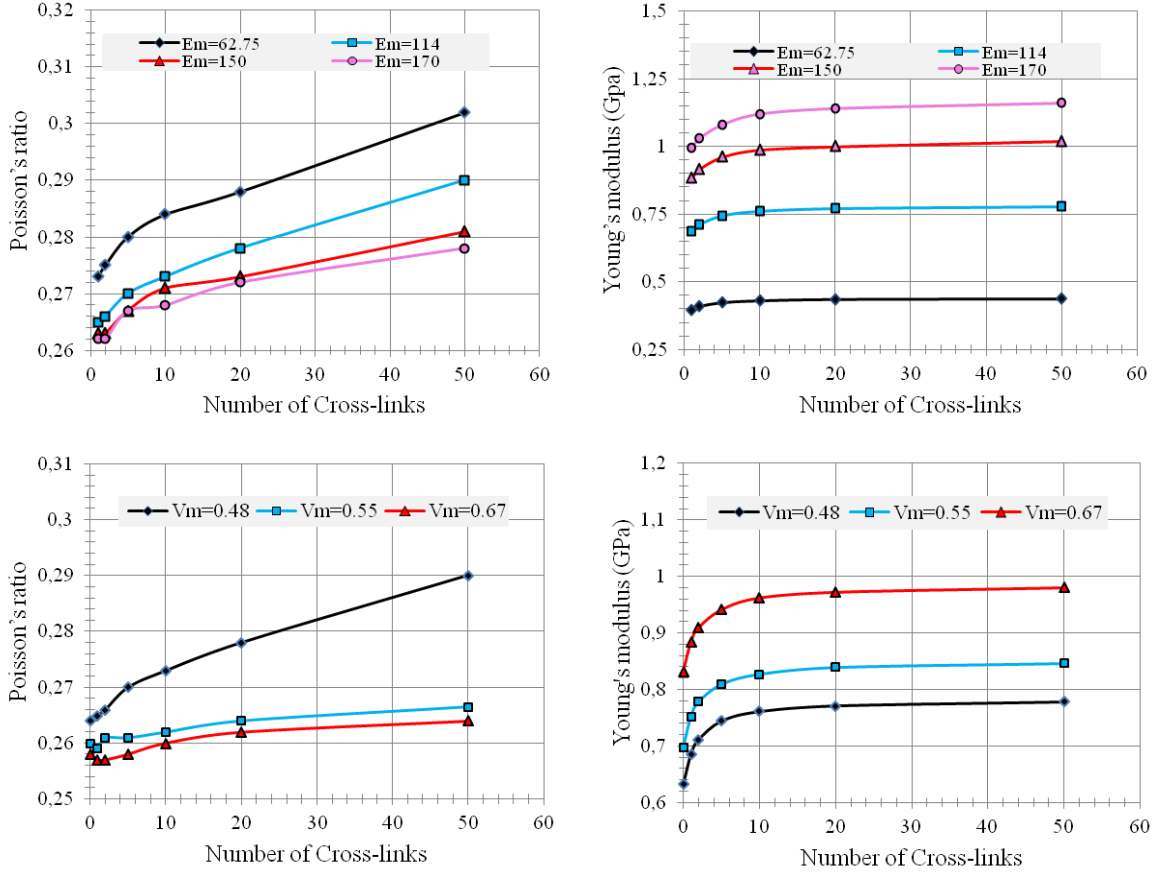

**Figure 4.** 3D FE simulation of collagen microfibril results: (a) Effect of Young's modulus of mineral and cross-link number. (b) Effect of volume fraction of mineral and cross-link number on elastic properties [2].

**b) Neural network modeling**

In this work an in-house NN algorithm developed by Hambli et al. [66] called Neuromod was used. This algorithm uses the total gradient method for neural networks construction and BP algorithm for training.

Input signals cumulated in the neuron block are activated by a nonlinear function given by:

$$f(v_m) = 1/(1 + e^{(-\beta v_m)}) \qquad (10)$$

Where β is a parameter defining the slope of the function.

This parameter which represents also the learning rate is adjusted arbitrary. In fact, we change this value many times from 0.001 to 0.1 in order to obtain the good combination which leads to the best fitting. It has a great influence on the NN performance. That's why; once we obtain a good fitting we kept the corresponding value all over our study. The good fitting in our case was obtained for β = 0.01.

The finite element modeling results of the mineralised collagen microfibril, presented in the previous section (Figure 4) were used for networks training.

**Result of NN training**

Because the FE calculation takes a long computing time and needs a lot of knowledge to adjust the numerical model, a meta-model based on NN was built and trained in order to substitute the FE calculation at nano-scale prediction step. The advantage of this NN meta-model that it is a black box and it doesn't need any prerequisites for model adjustments. Moreover it gives instantaneous response. But first of all, we have to check its accuracy compared to FE method. That s why, the error calculated according to equation below when was comparing the two responses (FE and NN).

$$error = \frac{Var^{FE} - Var^{NN}}{Var^{FE}} \quad (11)$$

Where, *Var* represents the Poisson ratio or the Young Modulus.

The maximum error doesn't exceed 0.001. As it can be shown from figure 5 that our NN meta-model gives good results. So, it was incorporated in the overall loop (Figure2).

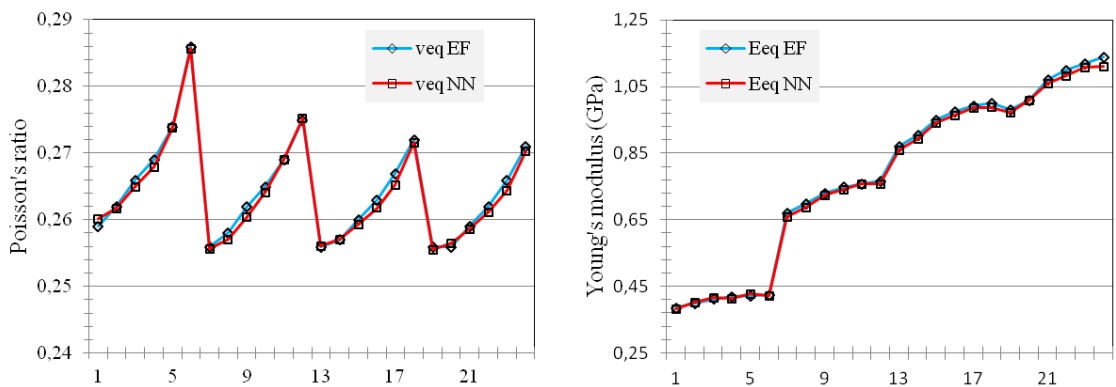

**Figure 5.** Validation of NN prediction

The results obtained by the NN method in step (1) were introduced as inputs to step (2) in order to calculate the elastic properties of mineralised collagen fibrils using the homogenisation equations (equations 4 and 11).

### 2.4.3. Step 2: Mineralised collagen fibril

To the best of our knowledge, in all earlier multiscale modeling studies, the mineralised collagen is considered to be the smallest component in cortical bone and there is no level scale between collagen one and fibrils. In the current work, a different point of view is proposed: a new level scale "mineralized collagen microfibril" is considered between bone elementary constituents (collagen and mineral) and mineralized collagen fibril. In the current investigation, microfibril refers to a composite material which consists of a mineral matrix reinforced by TC molecules connected by cross-links and modeled by NN and FE methods. The mineralised collagen fibril is modeled in a different manner to that proposed by Martínez-Reina et al.[22] i.e. the fibril scale is considered to be a composite in which the collagen microfibrils (matrix) embed platelet-shaped crystals (inclusion), which are assumed to be periodically distributed along the long axes of collagen fibrils. Nemat-Nasser and Hori [67] proposed a method to calculate the effective stiffness tensor of the composite with periodically distributed inclusions, as follows:

$$C_f = C_{mf}\left\{1 - \phi_{hp}\left[(C_{mf} - C_{hw})^{-1}C_{mf} - P_{p,mf}\right]^{-1}\right\} \quad (12)$$

where $C_f$ is the effective stiffness tensor of the collagen mineralised fibril; $C_{hw}$ is the stiffness tensor of the mineral-water composite obtained in step 0 and $C_{mf}$ is the mineralised collagen microfibril tensor. 1 is the identity tensor. $\phi_{hp}$ is the volumetric fraction of HA crystals in the composite, $P_{p,mf}$ corresponds to a periodic tensor operator for platelet-shaped inclusions of mineral crystals in a collagen microfibril matrix.

### 2.4.4. Step 3: Mineralised collagen fiber

The collagen fiber is considered to be a composite with a periodicity of cylindrical collagen fibrils that are embedded in a matrix of mineral-water composite. The effective stiffness tensor of the fiber $C_F$, is estimated in the same way that was used for the Fibril which was calculated in the previous step [22, 67, 75, 76]:

$$C_F = C'_{hw}\left\{1 - \phi_{fibril}\left[(C'_{hw} - C_f)^{-1}C'_{hw} - P_{c,hw}\right]^{-1}\right\} \quad (13)$$

ϕ$_{fibril}$ is the volumetric fraction of fibrils within the fiber. P$_{c,hw}$, is a periodic operator, corresponding to infinite cylindrical periodic inclusions of fibrils embedded in a matrix of extrafibrillar mineral.

### 2.4.5. Step 4: Single lamella

Similar to the higher scale modeling, the lamella stiffness tensor C$_l$ is estimated under the assumption that the lamella is a composite material with a matrix of extrafibrillar mineral-water composite and periodically distributed cylindrical inclusions of fibers [67]:

$$C_l = C'_{hw} \left\{ 1 - \phi_F \left[ (C'_{hw} - C_F)^{-1} C'_{hw} - P_{c,hw} \right]^{-1} \right\} \quad (14)$$

where ϕ$_F$ is the volumetric fraction of fibers. All collagen fibers in a lamella are assumed to be aligned in the longitudinal direction, and this type of bone is called a parallel fibered bone.

### 2.4.6. Step 5: Single lamella with canaliculi

In order to model the lamella level, we considered that the first pores are the canaliculi. The voids are infinite cylindrical periodically distributed inclusions with null stiffness [22]. The stiffness tensor of a lamella with canaliculi C$_{can}$ can be expressed by the following equation [67]:

$$C_{can} = C_l \left\{ 1 - P_{can} [1 - P_{c,l}]^{-1} \right\} \quad (15)$$

P$_{can}$ is the canalicular porosity, P$_{c,l}$ is the periodic operator for infinite cylindrical inclusions in a matrix of lamellar tissue.

### 2.4.7. Step 6: Laminate with LRC lamellae

The stiffness tensor of a laminate C$_{LRC}$ composed of the three layers in the three different directions of canaliculi (longitudinal (L); radial (R) and circumferential (C)) as sorted by Beno et al. [77] is calculated by using equations (16-17) [78]. Equation (15) is applied when i and j are 1, 2, 3 or 6

$$(C_{LRC})_{ij} = \sum_{k=1}^{3} V^k \left[ (C_{can}^k)_{ij} - \frac{(C_{can}^k)_{i3}(C_{can}^k)_{j3}}{(C_{can}^k)_{33}} + \frac{(C_{can}^k)_{i3} \sum_{\alpha=1}^{3} \frac{V^\alpha (C_{can}^\alpha)_{j3}}{(C_{can}^\alpha)_{33}}}{(C_{can}^k)_{33} \sum_{\alpha=1}^{3} \frac{V^\alpha}{(C_{can}^\alpha)_{33}}} \right] \quad (16)$$

Where $V^k$ is the volume fraction of the layer, and indices k = 1, 2, 3 correspond to L, R, C respectively.

In the case of i=1, 2, 3 or 6 and j=4 or 5, $(C_{LRC})_{ij} = (C_{LRC})_{ji} = 0$, and when i and j are 4 or 5 the components will be estimated by using equation (17):

$$(C_{LRC})_{ij} = \frac{\sum_{k=1}^{3} \frac{V^k}{\Delta_k}(C_{can}^k)_{ij}}{\sum_{k=1}^{3}\sum_{\alpha=1}^{3} \frac{V^k V^\alpha}{\Delta_k \Delta_\alpha} \Gamma_{k\alpha}} \quad (17)$$

where:

$$\begin{cases} \Delta_k = (C_{can}^k)_{44}(C_{can}^k)_{55} - (C_{can}^k)_{45}(C_{can}^k)_{54} \\ \Gamma_{k\alpha} = (C_{can}^k)_{44}(C_{can}^\alpha)_{55} - (C_{can}^k)_{45}(C_{can}^\alpha)_{54} \end{cases} \quad (18)$$

### 2.4.8. Step 7: Lamellae with lacunae

In this level, lacunar porosity is added in the lamellae under the assumption previously proposed by Martínez-Reina et al. [22]: Lacunae are periodically distributed ellipsoids.
The stiffness tensor of lamellae with lacunae $C_{lac}$, is estimated by equation (19):

$$C_{lac} = C_{LRC} \left\{ 1 - P_{lac}[1 - P_{e,LRC}]^{-1} \right\} \quad (19)$$

$P_{lac}$ is the lacunar porosity and $P_{e,LRC}$ is the periodic operator for ellipsoidal inclusions in a matrix of laminate LRC. All periodic operator tensor is calculated by Nemat-Nasser and Hori [67] method as a function of inclusion form (Ellipsoida, Cuboidal, Cylindrical shape). A detail is given in the appendix of the paper.

### 2.4.9. Step 8: single osteon

The effective stiffness tensor of the single osteon $C_{TrI}$ is estimated by the method of Yoon et al. [77]. This method consists in constructing upper and lower bounds of the effective transversely isotropic elastic constants using the known orthotropic values. The upper and lower bounds are usually very close, making it possible to use the mean value as an estimate of the transversely isotropic elastic constants of the osteon. The properties of a single drained osteon $C_{TrI}$, where the subscripts stands for transversely isotropic [22], can be estimated by symmetrization.

### 2.4.10. Step 9: Superposition of osteons

Martínez's equation is applied to estimate the effective stiffness tensor of osteons $C_{SCS}$ (newly formed and old osteons) [22].

$$C_{SCS} = \left\{\phi_a C_{TrI}^n [1 + \mathbb{P}_{cyl}(C_{TrI}^n - C_{SCS})]^{-1} + (1 - \phi_a) C_{TrI}^o [1 + \mathbb{P}_{cyl}(C_{TrI}^o - C_{SCS})]^{-1}\right\} \cdot \left\{\phi_a [1 + \mathbb{P}_{cyl}(C_{TrI}^n - C_{SCS})]^{-1} + (1 - \phi_a)[1 + \mathbb{P}_{cyl}(C_{TrI}^o - C_{SCS})]^{-1}\right\}^{-1} \quad (20)$$

where $C_{TrI}^n$ and $C_{TrI}^o$ are the stiffness tensors of newly formed osteons and old osteons respectively, and $\phi_a$ is the volume fraction of newly formed osteons. $\mathbb{P}_{cyl}$ is the Hill polarization tensor for cylindrical inclusions related to the Eshelby tensor $S_{cyl}$ and the stiffness tensor of the matrix by using the following equation [79]:

$$S_{cyl} = \mathbb{P}_{cyl} C_{TrI}^n \quad (21)$$

$C_{SCS}$ is solved iteratively by an initial value of $C_{SCS} = C_{TrI}^o$.

### 2.4.11. Step 10: Drained cortical tissue

The interpretation by Benveniste [80] of the Mori-Tanaka approach for void inclusions is adopted in the current investigation to calculate the stiffness tensors of cortical tissue.

$$C_d = C_{TrI} \left[1 + \frac{P_{vas}}{1 - P_{vas}} (1 - S_{cyl})^{-1}\right]^{-1} \quad (22)$$

where $P_{vas}$ is the vascular porosity.

**Model parameters**

As in any other composite material, the mechanical properties and volume fraction of bone components play an important role in its overall behavior [21]. A wide range of values for the mechanical properties of collagen and mineral has been reported in the literature. Different values of volume fraction have been used for the mineral crystals, varying from 32 to 52% [6, 81]. Some values of the mineral volume fraction reported in the literature are listed in Table 1.

| Reference | *Mineral volume fractions (%)* |
|---|---|
| Lees , 1987[24] | 45 |
| Sasaki et al. 1991[82] | 50 |
| Wagner and Weiner 1992[83] | 35 |
| Jager and Fratzl, 2000[84] | 43 |
| Kotha and Guzelso, 2000[85] | 50 |
| Ji and Gao, 2006[86] | 45 |
| Nikolov and Raabe, 2008[81] | 52 |

**Table 1.** Mineral volume fractions reported in the literature

The purpose of the current study is to introduce the microfibril scale into Martin's Model in order to investigate its effect on the next higher scale (mineralised collagen fibril). To achieve this, the mechanical properties and volume fraction values listed in tables 2 and 3 were assigned to generate our new model. The results obtained by our model are compared to those obtained by other previous models. The associated Young's modulus is taken as 0.138 GPa to match water's bulk modulus, 2.3 GPa [21].

| **Material** | **Elastic modulus (GPa)** | **Poisson's ratio** |
|---|---|---|
| Collagen | 2.5 | 0.28 |
| HA crystals | 120 | 0.23 |
| Water | 0.138 | 0.49 |

**Table 2.** Properties of bone components selected in this work

| *Value of volume fraction (%)* | **Value** |
|---|---|
| Water in the mineral-water $\phi_{wh}$ | 0.123 |
| Water in the mineral-water $\phi'_{wh}$ | 0.42 |
| HA crystals within the microfibril $\phi_{hp}$ | 0.238 |
| Fibrils within the fiber $\phi_f$ | 0.58 |
| Fibers within the lamella $\phi_F$ | 0.866 |
| Newly formed osteons $\phi_a$ | 0.66 |
| Canalicular porosity $P_{can}$ | 0.0418 |
| Lacunar porosity $P_{lac}$ | 0.0082 |
| vascular porosity $P_{vas}$ | 0.04 |

**Table 3.** Properties of bone components selected in this work

For simplicity, all the components are assumed to have a linear elastic and isotropic behavior. In the present paper we seek to present and to validate our proposed multiscale approach. At this stage of our work we will not study the variation in mechanical properties of the bone constituents and the number of cross-links. The number of cross-links selected in this work is N = 0.

## 3. Results and discussion

This study aims to investigate the influence of the existence of microfibrils on the mechanical properties of bone. The main objective is to propose a multiscale approach. The proposed model is tested with a well-defined configuration in order to validate our approach. A study of the variation in mechanical and geometrical parameters is beyond the scope of the present paper.

In the current work, the effect of the volume fraction of the different constituents of bone together with the porosity is studied as originally proposed by Yoon and Cowin [18, 19]. The main hypothesis of this work, which differentiates it from the approach of Martínez-Reina et al., Yoon and Cowin [18, 19, 22], is that the fibril is not the smallest scale of bone; an intermediate scale is introduced in this work i.e. the mineralised collagen microfibril [2]. The fibril scale is considered in this study as a composite in which the collagen microfibrils (matrix) embed platelet-shaped crystals (inclusion). The elastic tensor of this matrix is calculated by applying equation (12) using the same method as that previously applied by Martínez-Reina et al. [22] to calculate the tensor of the mineralised collagen fiber.

### 3.1. Nanostructural levels

In this section we consider the nanostructure scales i.e. microfibril, fibril and fiber. At the microfibril scale, the elastic tensor was calculated by the NN method, as mentioned in section 2.2. The cylindrical mineralised collagen microfibril with the stiffness tensor given by the NN is embedded unidirectionally in the extrafibrillar hydroxyapatite matrix. Similarly, the cylindrical collagen fibrils are grouped in a mineral matrix to form collagen fibers. The last two scales (fibril and fiber) were modeled using the homogenisation equations (11-12) under the assumption of Hamed et al. [21] i.e. the extrafibrillar mineral crystals comprise 25% of the total hydroxyapatite crystals. The elastic tensors of the nanostructural levels are given in table 4.

| Level | Elastic tensor (GPa) |
|---|---|
| **Mineralised collagen microfibril** | $\begin{pmatrix} 1.070 & 0.434 & 0.434 & 0 & 0 & 0 \\ 0.434 & 1.070 & 0.434 & 0 & 0 & 0 \\ 0.434 & 0.434 & 1.070 & 0 & 0 & 0 \\ 0 & 0 & 0 & 0.632 & 0 & 0 \\ 0 & 0 & 0 & 0 & 0.632 & 0 \\ 0 & 0 & 0 & 0 & 0 & 0.632 \end{pmatrix}$ |
| **Mineralised collagen fibril** | $\begin{pmatrix} 1.133 & 0.468 & 0.468 & 0 & 0 & 0 \\ 0.468 & 1.185 & 0.474 & 0 & 0 & 0 \\ 0.468 & 0.474 & 1.226 & 0 & 0 & 0 \\ 0 & 0 & 0 & 0.787 & 0 & 0 \\ 0 & 0 & 0 & 0 & 0.787 & 0 \\ 0 & 0 & 0 & 0 & 0 & 0.787 \end{pmatrix}$ |
| **Mineralised collagen fiber** | $\begin{pmatrix} 16.144 & 6.233 & 6.233 & 0 & 0 & 0 \\ 6.233 & 16.144 & 6.233 & 0 & 0 & 0 \\ 6.237 & 6.237 & 16.110 & 0 & 0 & 0 \\ 0 & 0 & 0 & 9.928 & 0 & 0 \\ 0 & 0 & 0 & 0 & 9.928 & 0 \\ 0 & 0 & 0 & 0 & 0 & 9.928 \end{pmatrix}$ |

**Table 4.** Elastic tenser of nanostructural levels

At the ultra-structure level, mineral and collagen are arranged into higher hierarchical levels to form microfibrils, fibrils and fibers [87]. The existence of sub-structures in collagen fibrils has long been a topic of debate. Recent studies suggest the presence of microfibrils in fibrils. Experimental work by Orgel, Fratzl and others prove that virtually all collagen-based tissues are organized into hierarchical structures, where the lowest hierarchical level consists of triple helical collagen molecules [88-90] and the multi-scale structure is defined as tropocollagen molecular-triple helical collagen molecules-fibrils-fibers. A longitudinal microfibrillar structure with a width of 4 - 8 nm was visualized [91-93]. Three dimensional image reconstructions of 36 nm-diameter corneal collagen fibrils also showed a 4 nm repeat in a transverse section, which was related to the microfibrillar structure [94]. Using X-ray diffraction culminating in an electron density map, Orgel et al. [90] suggested the presence of right-handed super-twisted microfibrillar structures in collagen fibrils.

Most models developed to predict the behavior of bone at the nanoscale used the continuum mechanics approach [6, 18, 19], while the nano-sized dimensions of bone components and their special arrangements motivated some researchers to use specific simulations. Buehler [95] used molecular dynamics (MD) simulation to study pure and mineralised collagen fibrils, whereas Bhowmik et al. [57] used MD to address the load carrying behavior of collagen in the proximity of HA. Dubey and Tomar [96] also analyzed the type-I collagen and HA arrangement using molecular dynamics. In the current study the mineralised collagen microfibril scale is introduced into the bone multiscale approach.

The concept and the mechanical properties were taken from our previous work Barkaoui et al., Barkaoui and Hambli, Hambli and Barkaoui [2, 53, 72] which is simulated for the first time by the finite element method. This scale has not been subjected to sufficient investigations. The elastic properties of microfibrils given in table 5 were estimated by Gautieri et al. [87] using MD simulation and by Van der Rijt et al., Shen et al. [97, 98] experimentally.

| Source | Type of analysis | Young's modulus (GPa) |
|---|---|---|
| Van der Rijt et al. 2006 [97] | AFM testing | 0.2-0.8 |
| Shen et al. 2008 [98] | MEMS stretching | 0.86 |
| Gautieri et al. 2011 [87] | Atomistic modeling | 1.2 |
| Present study | NN-EF modeling | 0.82 |

**Table 5.** Young's modulus of collagen mineralised microfibril

Recently Yang et al. [99] mentioned the existence of microfibrils in fibrils in their experimental investigation. Their study showed that the existence of microfibrils affects the mechanical behavior of fibrils: the estimated Young's modulus at different strain rates of native fibrils of 0.6 ± 0.2 GPa (table 6) supports the results of the current study at 0.9 GPa. However, when our results are compared with those which do not take this fibril substructure scale into consideration [18, 21], where a value of 3.4 GPa is found (table 6), the agreement is poor. This divergence in results provides support for our hypothesis of the important role of microfibrils.

| Levels | source | $E_1$(GPa) | $E_2$(GPa) | $E_3$(GPa) |
|---|---|---|---|---|
| **fibril** | Yoon and Cowin, (2008a) [18] | 3.4 | 6.7 | 7.7 |
| | Yang et al. (2012) [99] | 0.6 ± 0.2 | NA | NA |
| | Current study | 0.9 | 0.915 | 0.953 |
| **fibre** | Yoon and Cowin, (2008a) [18] | 11.3 | 13.8 | 16.9 |
| | Current study | 12.67 | 12.67 | 12.6 |

**Table 6.** Young's modulus of collagen mineralised fibril and fiber

*3.2. Microstructural levels*

To study the lamella scale, the different kinds of porosities (canalicules and lacunae) are taken into consideration in different orientations (LRC). These porosities were assumed to be ellipsoidal holes of dimensions $25 \times 10 \times 5$ μm [18,21] and Lacunar $P_{lac}$ and canalicular $P_{can}$ porosities of 0.8% and 4.18% respectively [22]. The osteon is assumed to be a cylinder 250 μm in diameter, 1 cm long and a Haversian canal with a diameter of 50 μm, giving a vascular porosity $P_{vas}$ approximately of 4% of the canal [21]. The stiffness tensors of the lamella at the different steps and of the single osteon were calculated and are shown in table 7.

| Level | Elastic tensor(GPa) |
|---|---|
| **Lamella** | $\begin{pmatrix} 16.08 & 6.21 & 6.21 & 0 & 0 & 0 \\ 6.21 & 16.08 & 6.21 & 0 & 0 & 0 \\ 6.22 & 6.22 & 16.07 & 0 & 0 & 0 \\ 0 & 0 & 0 & 9.88 & 0 & 0 \\ 0 & 0 & 0 & 0 & 9.88 & 0 \\ 0 & 0 & 0 & 0 & 0 & 9.88 \end{pmatrix}$ |
| **Lamellae with canaliculi** | $\begin{pmatrix} 16.18 & 6.24 & 6.24 & 0 & 0 & 0 \\ 6.24 & 16.18 & 6.24 & 0 & 0 & 0 \\ 6.25 & 6.25 & 16.14 & 0 & 0 & 0 \\ 0 & 0 & 0 & 9.96 & 0 & 0 \\ 0 & 0 & 0 & 0 & 9.96 & 0 \\ 0 & 0 & 0 & 0 & 0 & 9.96 \end{pmatrix}$ |
| **LRC Lamellae** | $\begin{pmatrix} 21.57 & 8.32 & 8.32 & 0 & 0 & 0 \\ 8.32 & 21.57 & 8.32 & 0 & 0 & 0 \\ 8.33 & 8.33 & 21.51 & 0 & 0 & 0 \\ 0 & 0 & 0 & 13.28 & 0 & 0 \\ 0 & 0 & 0 & 0 & 13.28 & 0 \\ 0 & 0 & 0 & 0 & 0 & 13.30 \end{pmatrix}$ |
| **Lamellae with lacunae** | $\begin{pmatrix} 19.44 & 8.24 & 8.25 & 0 & 0 & 0 \\ 8.24 & 20.34 & 8.25 & 0 & 0 & 0 \\ 8.25 & 8.25 & 32.2 & 0 & 0 & 0 \\ 0 & 0 & 0 & 14.02 & 0 & 0 \\ 0 & 0 & 0 & 0 & 12.33 & 0 \\ 0 & 0 & 0 & 0 & 0 & 12.54 \end{pmatrix}$ |

**Table 7.** Elastic tensor of microstructural levels

| Source | $E_1$(GPa) | $E_2$(GPa) | $E_3$(GPa) |
|---|---|---|---|
| Current study | 15.6 | 16.47 | 27.6 |
| Martínez-Reina et al. 2011 [22] | 17.2 | 19.7 | 22.0 |
| Yoon and Cowin 2008a [18] | 16.4 | 18.7 | 22.8 |
| Yoon and Cowin 2008b [19] | 16.9 | 19.0 | 22.3 |
| Rho et al. 2002 [100] | NA | NA | 21.8 ± 2.1 |
| Fan et al. 2002 [101] | 16.6 ± 1.5 | 17.0 ± 2.2 | 25.1 ± 2.1 |
| Zysset et al. 1999 [102] | 19.1 ± 5.4 | - | - |
| Hamed et al 2010 [21] | 17.91 | 11.88 | NA |
| Hoffler et al. 2000 [103] | 15.11 ± 2.2 | - | - |
| Wagner and Weiner 1998 [104] | 33,6 | 24.3 | NA |

**Table 8.** Young's modulus of collagen mineralised lamella

### 3.3. Mesostructural levels

In this simulation, a volume fraction of 66% for osteons [105] and a vascular porosity $P_{vas}$ of 4% [22] were chosen. The elastic tensors of osteons and cortical bone are given in table 9.

| Level | Elastic tensor(GPa) |
|---|---|
| Osteons | $\begin{pmatrix} 19.44 & 8.24 & 8.25 & 0 & 0 & 0 \\ 8.24 & 20.34 & 8.25 & 0 & 0 & 0 \\ 8.25 & 8.25 & 32.3 & 0 & 0 & 0 \\ 0 & 0 & 0 & 13.17 & 0 & 0 \\ 0 & 0 & 0 & 0 & 13.16 & 0 \\ 0 & 0 & 0 & 0 & 0 & 13.18 \end{pmatrix}$ |
| Cortical bone | $\begin{pmatrix} 21.96 & 8.24 & 10.19 & 0 & 0 & 0 \\ 8.24 & 23.00 & 8.25 & 0 & 0 & 0 \\ 10.19 & 8.25 & 26.04 & 0 & 0 & 0 \\ 0 & 0 & 0 & 6.89 & 0 & 0 \\ 0 & 0 & 0 & 0 & 6.12 & 0 \\ 0 & 0 & 0 & 0 & 0 & 6.45 \end{pmatrix}$ |

**Table 9** Elastic tensor of mesostructural levels

| Level | Source | E1( GPa) | E2( GPa) | E3( GPa) |
|---|---|---|---|---|
| **Osteon** | Current study | 17.15 | 18.12 | 29.7 |
| | Yoon and Cowin (2008b) [19] | 15.9 | 15.9 | 20.3 |
| | Ascenzi and Bonucci (1967)[106] | 21.1 ± 6.2 | | |
| **Cortical bone** | Current study | 16.91 | 18 | 19.40 |
| | Martínez-Reina et al (2011) [22] | 14.7 | 14.7 | 19.50 |
| | Yoon and Cowin (2008b) [19] | 20.3 | | |
| | Hamed et al. 2010 [21] | 11.84 | 11.84 | 18.69 |
| | Yoon et katz 1976 [107] (Ultrason) | 18.8±1.01 | 18.8±1.01 | 27.4±0.98 |
| | Terneur et al. 1999 [108] (Nanoindentation) | 16.58±0.32 | 16.58±0.32 | 23.45±0.21 |

**Table 10.** Young's modulus of single osteon and cortical bone

Tables 8 and 10 show, respectively, the longitudinal and transverse elastic moduli of the lamella, the single osteon and the cortical bone obtained using our model. These tables also give the selected experimental data available in the literature to enable comparison with our results. Our analytical results are in a reasonably good agreement with experiments. It should be noted that most of the studies report the average elastic modulus of cortical bone in the transverse direction.

The proposed approach is inspired from the work of Martínez-Reina et al. [22]. We followed their same assumptions of simplification as well as their selections of the different stages of modeling. We, however, proposed to add another sub level (microfibril) in order to keep a clear reference of comparison to represent the effect of this inserted new level on the mechanical behavior of the cortical bone.

Our definition of scales based on the approach by Martínez-Reina et al. [22] is not the only possible one. The transition between the different hierarchies of the nanoscale to macroscale is continuous rather than discrete in bone in vivo. While there is a general consensus concerning the classification of the main scales, there is a certain flexibility concerning the intermediate levels. For example Hamed et al. [21] minimized the number of scales; they did not consider the fiber level, between the nanoscale (mineralised collagen fibrils) and the sub-microscopic level (single lamella), in their investigation. Yang et al. [98], in contrast, suggest the existence of microfibrils.

In this study, we modeled the cortical bone as a hierarchical material and predicted its effective elastic moduli. Our analysis involved the bottom-up approach, starting with the nanostructural level (mineralised collagen microfibril level) and moving up to the mesoscale level (cortical bone level). The selection of the scales was not unique. In the analysis, we used the models of micromechanics and composite laminate theory. The contribution of this work involves the multiscale modeling of bone from nano to mesoscale levels. The contribution is also to identify of the challenges involved in modeling bone at each structural scale using deferent methods. In this paper, the method developed by Martínez-Reina et al. [22] to estimate the elastic properties of cortical bone scales levels has been slightly modified to include the mineralised collagen microfibril scale using finite element method [72] and NN method [3]. It has also been used to study the effect of this sub-structure in the upper scale behavior.

**Acknowledgements**

This work was supported by the French National Research Agency (ANR) through the TecSan program (Project MoDos, n°ANR-09-TECS-018) and part of the Fractos project supported by the Region Centre (France). The authors gratefully acknowledge Mr Bettamer for his help.

**Appendix**

**Periodic operator tensor**

The fourth rank tensor of the periodic operator for isotropic material is given by Nemat-Nasser and Hori [67]:

$$P = \sum_{\xi}{}' \phi g(-\xi)\, g(\xi) F\hat{P}(\xi)$$

Where the prime on the summation indicates that $\xi = 0$ is excluded. $g(\xi)$ is called the g-integral. The geometry of a cavity is represented by the g-integral, which is defined as the volume integral of $\exp(i\xi.x)$ over the cavity. The newly defined variable $\xi$ in the equation is given by $\xi = \frac{n_i \pi}{a_i}$, where $n_i$ is the number of unit cells in the i-direction. Note that i can be replaced with x, y, z for the x-, y-, and z-directions, respectively, and the domain of a unit cell is given by: $U = \{x; -a_i < x_i < a_i (i = x, y, z)\}$

The term $F\hat{P}(\xi)$ is the second rank tensor in 6 D and it can be determined from the fourth rank tensor $FP(\xi)$ in 3D (For details, wee [67]:

$$FP(\xi) = \text{sym}\{\xi \otimes (\xi.C^m.\xi)^{-1} \otimes \xi\} : C^m$$

$C^m$ is the is the fourth rank elasticity tensor of the bone matrix. Once the term $(\xi \cdot C \cdot \xi)$ has been expanded in the index notation as:

$$\begin{aligned}(\xi_i C_{ijkl} \xi_l) = &\ \xi_1 C_{1jk1} \xi_1 + \xi_1 C_{1jk2} \xi_2 + \xi_1 C_{1jk3} \xi_3 \\ &+ \xi_2 C_{2jk1} \xi_1 + \xi_2 C_{2jk2} \xi_2 + \xi_2 C_{2jk3} \xi_3 \\ &+ \xi_3 C_{3jk1} \xi_1 + \xi_3 C_{3jk2} \xi_2 + \xi_3 C_{3jk3} \xi_3 \end{aligned}$$

The inverse of the second rank tensor $(\xi_i C_{ijkl} \xi_l)$ is calculated. By the tensor operation, another fourth rank tensor $\left(\xi_i \left(\xi_p C_{pjkq} \xi_q\right)^{-1} \xi_{lp}\right)$ is formed. Then the quantity FP($\xi$) is obtained by the contraction of this result with the fourth rank elasticity tensor of bone matrix $C^m$, and converted into the Kelvin second rank tensor $\widehat{FP}(\xi)$. Note that the summation in the above equation with respect to $\xi$ (or $n_x$, $n_y$, and $n_z$) is generally from 1 to infinity, but in this study, the summation is performed from one to ±50 because Nemat-Nasser et al. [67] showed that the summation up to ±40 is only 0.7% less accurate than the summation up to ±50.

The g-integral g($\xi$) is calculated as a function of the inclusion shape. Three cases are distinguished: ellipsoidal, cubical and cylindrical.

**1-Ellipsoidal shape**

The g-integral g($\xi$) for the ellipsoidal inclusion is given by Nemat-Nasser and Hori [67]:

$$g(\xi) = \frac{3}{\eta}(\sin\eta - \eta\cos\eta)$$

and

$$\eta = 2 \times \pi \times \left(\frac{3 \times \phi}{4 \times \pi \times (b_y/b_x) \times (b_z/b_x)}\right) \times \left[(n_x)^2 + \left(n_y \frac{b_y}{b_x}\right)^2 + \left(n_z \frac{b_z}{b_x}\right)^2\right]^{1/2}$$

x, y and z indicate three perpendicular directions. The number of cavities in the x, y, and z directions are denoted by $n_x$, $n_y$ and $n_z$. Respectively the length of the principal axes of an ellipsoidal shape representing a lacuna are denoted by $b_x$, $b_y$ and $b_z$. For the type L osteon, the approximate dimensions of a lacuna are $b_x = 5\mu m$, $b_y = 10\mu m$ and $b_z = 25\mu m$

**2-Cuboidal shape**

The g-integral for the platelet shaped (or cubical) inclusion is given by Nemat-Nasser and Hori [67]

$$g(\xi) = \frac{\sin L_x \sin L_y \sin L_z}{L_x L_y L_z}$$

where $L_i = \xi_i l_i$

The subscript i is replaced by x, y, and z to indicate three perpendicular directions, and $l_i$ is the dimension of the platelet mineral crystals, i.e., $l_x = 3nm, l_y = 25\ nm, l_z = 50nm$.

**3- Cylindrical shape**

The g-integral, $g(\xi)$ for the circular cylindrical shape of the mineralised collagen fibrils, is given by Nemat-Nasser and Hori [67]

$$g(\xi) = \begin{cases} \frac{2}{B} J_1(B) & \text{if } n_z = 0 \\ 0 & \text{if } n_z \neq 0 \end{cases}$$

$J_1$ is the Bessel function of the first kind, and

$$B = 2\pi(n_x^2 + n_y^2)^{1/2} \sqrt{\frac{\phi_F}{\pi}}$$

$n_x$ and $n_y$ are the number of cylindrical inclusions in the x and y directions, respectively. $\phi_F$ is the volumetric fraction of fibers.

# References:

[1] Sergey V. Dorozhkin, Nanosized and nanocrystalline calcium orthophosphates, ActaBiomaterialia 6 (2010) 715–734.

[2] Barkaoui A, Hambli R. Finite element 3D modeling of Mechanical Behaviour of Mineralized collagen Microfibril, J ApplBiomaterBiomech 9(3) (2011) 207 – 213.

[3] Hambli R., Katerchi H. and Benhamou C.L., Multiscale methodology for bone remodelling simulation using coupled finite element and neural network computation, Biomechanics and Modeling in Mechanobiology ,10(1)(2010), 133-145.

[4] Katz, J.L., Anisotropy of Young's modulus of bone, Nature 283 (1980), 106-107.

[5] Katz, J.L., Meunier, A. The elastic anisotropy of bone. J. Biomech. 20(1987), 1063–1070 .

[6] Fritsch, A., Hellmich, C. Universal microstructural patterns in cortical and trabecular, extracellular and extravascular bone materials: micromechanics-based prediction of anisotropic elasticity. J. Theor. Biol. 244(2007)597–620

[7] Feng, L. Multi-scale characterization of swine femoral cortical bone and long bone defect repair by regeneration. Ph.D. dissertation, University of Illinois at Urbana-Champaign (2010)

[8] Deuerling JM, Yue W, Espinoza Orías AA, Roeder RK, Specimen-specific multi-scale model for the anisotropic elastic constants of human cortical bone. J Biomech 42 (2009)2061–2067

[9] Dong XN, Guo XE, Prediction of cortical bone elastic constants by a two-level micromechanical model using a generalized selfconsistent method. J BiomechEng 128(2006)309–316

[10] Ghanbari J, Naghdabadi R. Nonlinear hierarchical multiscale modeling of cortical bone considering its nanoscale microstructure. J Biomech 42: (2009) 1560–1565.

[11] Porter D, Pragmatic multiscalemodelling of bone as a natural hybrid nanocomposite. Mat ScieEng A 365(2004)38–45.

[12] Kotha SP, Guzelsu N, Tensile behavior of cortical bone: dependence of organic matrix material properties on bone mineral content. J Biomech 40(2007)36–45.

[13] Sevostianov I, Kachanov M, Impact of the porous microstructureon the overall elastic properties of the osteonal cortical bone.J Biomech 33(2000)881–888.

[14] Fritsch A, Hellmich C, Dormieux L., Ductile sliding between mineral crystals followed by rupture of collagen crosslinks: experimentally supported micromechanical explanation of bone strength, J Theor Biol. (2009)260(2):230-52.

[15] Brynk T, Hellmich C, Fritsch A, Zysset P, Eberhardsteiner J., Experimental poromechanics of trabecular bone strength: role of Terzaghi's effective stress and of tissue level stress fluctuations, J Biomech. (2011)44(3):501-8.

[16] E. Budyn and T. Hoc, Analysis of micro fracture in human Haversian cortical bone under transverse tension using extended physical imaging, Int. J. Numer. Meth. Engng (2010) 82:940–965

[17] J. Jonvaux, T. Hoc and É. Budyn, Analysis of micro fracture in human Haversian cortical bone under compression, Int. J. Numer. Meth. Biomed. Engng. (2012) 28:974–998.

[18] Yoon YJ,Cowin SC, The estimated elastic constants for a single bone osteonal lamella. Biomech Model Mechanobiol 7:(2008a)1–11.

[19] Yoon YJ, Cowin SC, An estimate of anisotropic poroelastic constants of an osteon. Biomech Model Mechanobiol 7(2008b)13–26.

[20] Sansalone. V, ThibaultLemaire, Salah Naili, Variational homogenization for modeling fibrillar structures in bone. Mechanics Research Communications 36 (2009) 265–273.

[21] Elham Hamed, Yikhan Lee, IwonaJasiuk, Multiscale modeling of elastic properties of cortical bone, ActaMech 213(2010)131–154

[22] J. Martínez-Reina,J. Domínguez, J. M. García-Aznar, Effect of porosity and mineral content on the elastic constantsof cortical bone: a multiscale approach , Biomech Model Mechanobiol 10(2011) 309–322.

[23] Landis WJ, HodgensKJ, Arena J, Song MJ, McEwen BF. Structural relations between collagen and mineral in bone by high voltage electron microscopic tomography. Microscopy Research and Technique, 33(2)(1996) 192-202.

[24] Lees, S.,. Considerations regarding the structure of the mammalian miner- alized osteoid from viewpoint of the generalized packing model. Connect. Tissue Res 16(1987), 281–303.

[25] Buehler, M.J., Nanomechanics of collagen fibrils under varying cross-link densities: atomistic and continuum studies. J. Mech. Behaviour Biomed.Mater. 1 (1)(2008), 59–67.

[26] Uzel, S.G.M., Buehler, M.J., 2011. Molecular structure, mechanical behaviour and failure mechanism of the C-terminal cross-link domain in type I collagen. J. Mechan. Behav.Biomed.Mater. 4, 153–161.

[27] Posner AS. Crystal chemistry of bone mineral.Phys.Rev. 49(1969) 760-792.

[28] Marc D. Grynpas, Laurence C. Bonar, and Melvin J. Glimcher, Failure to Detect an Amorphous Calcium-Phosphate Solid Phase in Bone Mineral: A Radial Distribution Function Study, Calcif Tissue Int (1984) 36:291-301

[29] Rubin MA, Jasiuk I, Taylor J, Rubin J, Ganey T, Apkarian RP. TEM analysis of the nanostructure of normal and osteoporotic human trabecular bone. Bone 33(2003)270–282.

[30] Pidaparti, R.M., Chandran, A., Takano, Y., Turner, C.H.,. Bone mineral lies mainly outside collagen fibrils: predictions of a composite model for osteonal bone. J. Biomech 29(1996)909–916.

[31] C., Hellmich and F.-J., Ulm, Average hydroxyapatite concentration is uniform in the extracollagenous ultrastructure of mineralized tissues: evidence at the 1–10-μm scal, Biomech Model Mechanobiol 2(2003)(1)21-36.

[32] Vashishth, D.,. The role of collagen matrix in skeletal fragility.Curr.Osteoporos. Rep. 5(2007) 62–66.

[33] Matthew J. Olszta , Xingguo Cheng , Sang SooJee , Rajendra Kumar , Yi-Yeoun Kim , Michael J. Kaufman , Elliot P. Douglas , Laurie B. Gower, Bone structure and formation: A new perspective, Materials Science and Engineering R 58 (2007) 77–116.

[34] Yan Liu , Young-Kyung Kim , Lin Dai , Nan Li , Sara O. Khan , David H. Pashley , Franklin R. Tay, Hierarchical and non-hierarchical mineralisation of collagen, Biomaterials 32 (2011) 1291-1300.

[35] Ch. Hellmich, F.-J. Ulm , Are mineralized tissues open crystal foams reinforced by crosslinked collagen?–some energy arguments, Journal of Biomechanics 35(9) (2002),1199‑1212

[36] V. Benezra Rosena, L.W. Hobbsa, M. Spectorb, The ultrastructure of anorganic bovine bone and selected synthetic hyroxyapatites used as bone graft substitute materials, Biomaterials 23 (2002) 921–928.

[37] Benjamin Alexander, Tyrone L. Daulton, Guy M. Genin, Justin Lipner, Jill D. Pasteris, Brigitte Wopenka and Stavros Thomopoulos, The nanometre-scale physiology of bone: steric modelling and scanning transmission electron microscopy of collagen–mineral structure, J. R. Soc. Interface. 9 (2012) 1774-1786.

[38] John D. Currey, The structure and mechanics of bone, J Mater Sci (2012) 47:41–54

[39] Gelse, K.,Poschl,E.,Aigner,T.,. Collagens-structure, function, and biosynthesis. Adv.Drug Deliv.Rev.55(12) (2003),1531–1546.

[40] Knott, L., Bailey, A.J.,. Collagen cross-links in mineralizing tissues: a review of their chemistry, function, and clinical relevance. Bone 22 (3)(1998)181–187.

[41] Sasaki, N., Odajima, S.,.Elongation mechanism of collagen fibrils and force– strain relations of tendon at each level of structural hierarchy. J. Biomech 29 (9)(1996)1131–1136.

[42] Gupta, H.S., Seto, J., Krauss, S., Boesecke, P., Screen, H.R.C., In situ multi-level analysis of viscoelastic deformation mechanisms in tendon collagen. J. Struct. Biol. 169 (2)( 2010), 183–191

[43] Eyre, D.R., Weis, M.A., Wu, J.J., Advances in collagen crosslink analysis. Methods 45 (1)(2008)65–74.

[44] Bailey, A.J., Molecular mechanisms of ageing in connective tissues.Mech. Ageing Dev. 122 (7)(2001),735–755.

[45] Wu, P., Koharski, C., Nonnenmann, H., Vashishth, D., Loading on non- enzymatically glycated and damaged bone results in an instantaneous fracture. Trans. Orthopaedic Res. Soc. 28(2003), 404.

[46] Vashishth, D., Wu, P., Gibson, G., Age-related loss in bone toughness is explained by non-enzymatic glycation of collagen. Trans. Orthopaedic Res. Soc., 29(2004).

[47] Boxberger, J., Vashishth, D., Nonenzymaticglycation affects bone fracture by modifying creep and inelastic properties of collagen. Trans. Orthopaedic Res. Soc. 29(2004), 0491.

[48] Catanese J., Bank R., Tekoppele J., Keaveny T., Increased crosslinking by non-enzymatic glycation reduces the ductility of bone and bone collagen. In: Proceedings of the Am. Soc. Mech. Eng. Bioeng. Conf. 42: (1999) 267–268.

[49] Allen, M.R., Gineyts, E., Leeming, D.J., Burr, D.B., Delmas, P.D., Bispho- sphonates alter trabecular bone collagen cross-linking and isomerization in beagle dog vertebra. Osteoporos. Int. 19 (3)(2007), 329–337.

[50] Tang, S., Bank, R., Tekoppele, J., Keaveny, T.,.Nonenzymaticglycation causes loss of toughening mechanisms in human cancellous bone. Trans. Orthopaedic Res. Soc. 30(2005).

[51] Siegmund, T., Allen, M.R., Burr, D.B., 2008. Failure of mineralized collagen fibrils: Modeling the role of collagen cross-linking. J. Biomech. 41, 1427–1435.

[52] Saito, M., Marumo, K., Collagen cross-links as a determinant of bone quality: a possible explanation for bone fragility in aging, osteoporosis, and diabetes mellitus. Osteoporos. Int. 21 (2)(2009),195–214.

[53] Barkaoui A, Bettamer A, Hambli R. Failure of mineralized collagen microfibrils using finite element simulation coupled to mechanical quasi-brittle damage, Procedia Engineering 10 (2011) 3185–3190.

[54] Barkaoui A and Hambli R (2013): Nanomechanical properties of mineralised collagen microfibrils based on finite elements method: biomechanical role of cross-links, Computer Methods in Biomechanics and Biomedical Engineering, DOI:10.1080/10255842.2012.758255.

[55] Currey, J.D., Role of collagen and other organics in the mechanical properties of bone.Osteoporos Int. 14(2003), S29–S36.

[56] Wilson, E., Awonusi, A., Morris, M., Kohn, D., Tecklenburg, M., Beck, L.,Three structural roles for water in bone observed by solid-state NMR.Biophys, J, 90(2006), 3722–3731.

[57] Bhowmik, R., Katti, K.S., Katti, D.R., Mechanics of molecular collagen is influenced by hydroxyapatite in natural bone. J. Mater. Sci. 42(2007) 8795–8803.

[58] Wang, X.D., Puram, S., The toughness of cortical bone and its relationship with age. Ann. Biomed Eng. 32(2004)123–135.

[59] Nyman, J.S., Roy, A., Shen, X.M., Acuna, R.L., Tyler, J.H., Wang, X.D., The influence of water removal on the strength and toughness of cortical bone. J. Biomech 39(2006),931–938.

[60] Sasaki, N., Enyo, A., Viscoelastic properties of bone as a function of water- content. J. Biomech. 28(1995), 809–815.

[61] Knowles, T.P., Fitzpatrick, A.W., Meehan, S., Mott, H.R., Vendruscolo, M., Dobson, C.M., Welland, M.E., Role of intermolecular forces in defining material properties of protein nanofibrils. Science 318(2007), 1900–1903.

[62] L. Eberhardsteiner, C. Hellmich & S. Scheiner (2013), Layered water in crystal interfaces as source for bone viscoelasticity: arguments from a multiscale approach, , Comp Meth Biomech Biomed EngDOI:10.1080/10255842.2012.670227.

[63] Topping BHV, Bahreininejad A,. Neural computing for structural mechanics. Saxe Coburg, (1992) UK

[64] JenkinsWM (An introduction to neural computing for the structural engineer. StructEng 75(3)1997) 38–41.

[65] Rafiq MY, Bugmann G, Easterbrook DJ, Neural network design for engineering applications. ComputStruct 79 (17) (2001) 1541–1552.

[66] Hambli R, Chamekh A, BelHadj Salah H, Real-time deformation of structure using finite element and neural networks in virtual reality applications. Finite Elem Anal Des 42(11)(2006) 985–991

[67] Nemat-Nasser S, Hori, M Micromechanics: overall properties ofheterogeneous materials, 2nd edn. Elsevier (1999), Amsterdam.

[68] Chou PC, Carleone J, Hsu CM, Elastic constants of layered media. J Compos Mater 6: (1972) 80–93.

[69] C. Morin, C. Hellmichn, P. Henits, Fibrillar structure and elasticity of hydrating collagen: A quantitative multiscale approach, Journal of Theoretical Biology 317 (2013) 384–393.

[70] Gong JK, Arnold JS, Cohn SH (1964). Composition of trabecular and cortical bone. Anat Rec 149: 325-332.

[71] Biltz, R. and Pellegrino, E. (1969). The chemical anatomy of bone. Journal of Bone
and Joint Surgery, 51-A(3):456 – 466.

[72] Hambli R, Barkaoui A, Physically based 3D finite element model of a single mineralized collagen microfibril, Journal of Theoretical Biology 301 (2012), 28–41

[73] Smith JW. Molecular Pattern in Native Collagen, Nature. 219(1968) 157-158.

[74] Fang Yuan, Stuart R. Stock, Dean R. Haeffner, Jonathan D. Almer, David C. Dunand ,L. Catherine Brinson, A new model to simulate the elastic properties of mineralizedcollagen fibril, Biomech Model Mechanobiol.10(2011) 147–160

[75] C. Hellmich and F. Ulm (2002). "Micromechanical Model for Ultrastructural Stiffness of Mineralized Tissues." J. Eng. Mech.,128(8), 898–908.

[76] C. Hellmich, J.F. Barthélémy, L. Dormieux, Mineral–collagen interactions in elasticity of bone ultrastructure – a continuum micromechanics approach, European Journal of Mechanics A/Solids 23 (2004) 783–810.

[77] Beno T, Yoon YJ, Cowin SC, Fritton S, Estimation of bone permeability using accurate microstructuralmeasurement. J Biomech 39: (2006) 2378–2387.

[78] Yoon YJ, Yang G, Cowin SC. Estimation of the effective transversely isotropic elastic constants of a material from known values of the material's orthotropic elastic constants. Biomech Model Mechanobiol 1(2002)83–93.

[79] Suvorov AP, Dvorak GJ (2002) Rate form of the Eshelby and Hill tensors. Int J Solids Struct 39:5659–5678.

[80] Benveniste Y, A new approach to the application of Mori-Tanaka's theory in composite materials.Mech Mater 6: (1987) 147–157

[81] Nikolov, S., Raabe, D.,. Hierarchical modeling of the elastic properties of bone at submicron scales: the role of extrafibrillar mineralization. Biophys. J. 94(2008) 4220–4232

[82] Sasaki, N., Ikawa, T., Fukuda, A.: Orientation of mineral in bovine bone and the anisotropic mechanical properties of plexiform bone. J. Biomech. 24(1991)57–61

[83] Wagner, H.D.,Weiner, S.: On the relationship between the microstructure of bone and its mechanical stiffness. J. Biomech. 25(1992)1311–1320

[84] Jager, I., Fratzl, P.: Mineralized collagen fibrils: a mechanical model with a staggered arrangement of mineral particles. Biophys. J. 79(2000)1737–1746

[85] Kotha, S.P., Guzelsu, N. The effects of interphase and bonding on the elastic modulus of bone: changes with age-related osteoporosis. Med. Eng. Phys. 22(2000) 575-585.

[86] Ji, B., Gao, H.: Elastic properties of nanocomposite structure of bone. Compos. Sci. Technol. 66(2006)1209–1215.

[87] Gautieri A, Vesentini S, Redaelli A, Buehler MJ, Hierarchical structure and nanomechanics of collagen microfibrils from the atomistic scale up, Nano Letters 9,11(2) (2011) 757-66.

[88] Fratzl, P., Collagen: Structure and Mechanics. Springer (New York) (2008).

[89] Orgel, J.P.R.O., A. Miller, T.C. Irving, R.F. Fischetti, A.P. Hammersley, and T.J. Wess, The In Situ Supramolecular Structure of Type I Collagen. Structure, 9(2001) 1061-1069.

[90] Orgel, J.P.R.O., T.C. Irving, A. Miller, and T.J. Wess, Microfibrillar structure of type I collagen in situ. P. Natl. Acad. Sci. USA,103 (24) (2006) 9001-9005.

[91] Raspanti M, Congiu T, Guizzardi S. Tapping-mode atomic force microscopy in fluid of hydrated extracellular matrix. Matrix Biol.20 (2001) 601-604.

[92] Habelitz S, Balooch M, Marshall SJ, Balooch G, Marshall GW. In situ force microscopy of partially demineralized human dentin collagen fibrils. J. Struct. Biol. 138 (2002),227-236.

[93] Baselt DR, Revel JP, Baldschwieler JD.Subfibrillar structure of type I collagen observed by atomic force microscopy. Biophys. J. 65(1993)2644-2655.

[94] Holmes DF, Gilpin CJ, Baldock C, Ziese U, Koster AJ, Kadler KE. Corneal collagen fibril structure in three dimensions: structural insights into fibril assembly, mechanical properties, and tissue organization. Proc. Natl. Acad. Sci. USA. 98 (2001) 7307-7312.

[95] Buehler, M.J.: Molecular nanomechanics of nascent bone: fibrillar toughening by mineralization. Nanotechnology 18(2007) 295102–295111

[96] Dubey, D.K., Tomar, V.: Microstructure dependent dynamic fracture analyses of trabecular bone based on nascent bone atomistic simulations. Mech. Res. Commun. 35(2008)24–31

[97] Van der Rijt, J.A.J., K.O. van der Werf, M.L. Bennink, P.J. Dijkstra, and J. Feijen, Micromechanical testing of individual collagen fibrils. Macromolecular Bioscience, 6(9) (2006) 697-702.

[98] Shen ZL, Dodge MR, Kahn H, Ballarini R, Eppell SJ. Stress-strain experiments on individual collagen fibrils. Biophysical Journal, 95(8)(2008) 3956-3963.

[99] L. Yang, K.O. van der Werf, P.J. Dijkstra, J. Feijen, M.L. Bennink, Micromechanical analysis of native and cross-linked collagen type I fibrils supports the existence of microfibrils? ; Journal of the mechanical behavior of biomedical materials (2012) 148-158.

[100] Rho JY, Zioupos P, Currey JD, Pharr GM Microstructural elasticity and regional heterogeneity in aging human bone examined by nano-indentation. J Biomech 35(2002)161–165

[101] Fan Z, Swadener JG, Rho JY, Roy ME, Pharr GM, Anisotropicproperties of human tibial cortical bone as measured by nanoindentation.J Orthop Res 20: (2002) 806–810

[102] Zysset, P.K., Guo, X.E.,Hoffler, C.E.,Moore, K.E.,Goldstein, S.A. Elastic modulus and hardness of cortical and trabecularbone lamellae measured by nanoindentation in the human femur. J. Biomech. 32(1999) 1005–1012).

[103] Hoffler, C.E., Moore, K.E., Kozloff, K., Zysset, P.K., Goldstein, S.A.: Age, gender, and bone lamellae elastic moduli.J. Orthop. Res. 18(2000) 432–437

[104] Weiner S,Wagner HD . The material bone: structure mechanical function relations. Ann Rev Mater Res 28(1998) 271–298.

[105] Cowin, S.C.: Bone Mechanics Handbook. CRC Press, Boca Raton (2001)

[106] Ascenzi, A., Bonucci, E.: The tensile properties of single osteons. Anat. Rec. 158(1967), 375–386.

[107] Yoon HS, Katz JL. Ultrasonic wave propagation in human cortical bone. II Measurements of elastic properties and microhardness, Journal of Biomechanics 9(1976) 459-464

[108] Turner CH, Rho J, Takano Y, Tsui TY, Pharr GM, The elastic properties of trabecular and cortical bone tissues are similar: results from two microscopic measurement techniques. Journal of Biomechanics 32 (1999) (4) 437-441.